\documentclass[10, conference]{IEEEtran}
\usepackage{amsmath,amssymb,amsfonts}
\usepackage{bm}
\usepackage{bbm}
\usepackage{graphicx}
\usepackage{cite}
\usepackage{epstopdf}
\usepackage{balance}
\usepackage{gensymb}
\usepackage{color}
\usepackage{lipsum}
\usepackage{float}
\usepackage{epstopdf}
\usepackage[mathcal]{eucal}
\usepackage{subfiles}
\usepackage[T1]{fontenc}
\usepackage{siunitx}
\usepackage{tabulary}
 \DeclareMathAlphabet\mathbfcal{OMS}{cmsy}{b}{n}
\usepackage{epsfig,graphics,graphicx,amssymb,amstext,amsmath,algorithm,algorithmic,wrapfig,multirow}
\usepackage{array}
\usepackage{adjustbox}
\usepackage[dvipsnames]{xcolor}
\usepackage{cite}
\usepackage{times}
\usepackage{placeins}
\usepackage{textcomp}
\usepackage[font=small]{caption}
\usepackage{xcolor}
\usepackage{tabularx}
\usepackage{bm}
\usepackage{enumerate}
\usepackage{tikz}
\usepackage{pgfplots}
\usepackage{epstopdf}
\usetikzlibrary{shapes,arrows}
\usepackage[utf8]{inputenc}
\usepackage{mathtools}
\usepackage{xcolor}
\usepackage{tikz}
\usepackage{multirow}
\usetikzlibrary{chains,arrows,calc,positioning}
\usepackage{amssymb}
\usepackage{pifont}
\usepackage{caption}
\usepackage{subcaption}
\usepackage[letterpaper, top=0.75in, bottom=1in, left=0.625in, right=0.625in]{geometry}
\usepackage[T1]{fontenc}
\usepackage[bookmarks=false]{hyperref}

\pgfplotsset{compat=1.16}
\definecolor{bittersweet}{rgb}{1.0, 0.44, 0.37}
\definecolor{glaucous}{rgb}{0.38, 0.51, 0.71}
\definecolor{gainsboro}{rgb}{0.86, 0.86, 0.86}
\definecolor{babyblueeyes}{rgb}{0.63, 0.79, 0.95}
\definecolor{silver}{rgb}{0.75, 0.75, 0.75}
\definecolor{neoncarrot}{rgb}{1.0, 0.64, 0.26}

\usepackage{soul}

\usepackage{breqn}
\usepackage{booktabs}
\usepackage{multirow}

\usepackage{enumitem}
\usepackage{tabulary}

\usepackage[normalem]{ulem}

\makeatletter

 \let\oldforeign@language\foreign@language
 \DeclareRobustCommand{\foreign@language}[1]{%
   \lowercase{\oldforeign@language{#1}}}

\usepackage{dblfloatfix}
\usepackage{makecell}
\usepackage{soul,xcolor}

\setstcolor{red}
\usepackage[bookmarks=false]{hyperref}
\makeatother

\IEEEoverridecommandlockouts

\begin{document}
\title{{Low-Complexity Learning-Based Beamforming for Ultra-Massive MIMO THz Communications}}
\author{\IEEEauthorblockN
{
Sourabh Solanki$^{\ddagger}$, Abuzar Babikir Mohammad Adam$^{\dagger}$, Chandan Kumar Sheemar$^{\dagger}$, Zaid Abdullah$^{\dagger}$,\\ Eva Lagunas$^{\dagger}$, George C. Alexandropoulos$^*$,  and
Symeon Chatzinotas$^{\dagger}$
}  
\\
$^{\ddagger}$Department of Electronics and Communication Engineering, NIT Warangal, India\\
$^{\dagger}$SnT, University of Luxembourg, Luxembourg
\\
 $^*$Department of Informatics and Telecommunications, National and Kapodistrian University of Athens, Greece\\
e-mails: \{ssolanki@nitw.ac.in\}\{abuzar.babikir, chandankumar.sheemar, zaid.abdullah\}@uni.lu,\\
\{eva.lagunas, symeon.chatzinotas\}@uni.lu,\{alexandg@di.uoa.gr\}
}
\maketitle

 \begin{abstract} 
Terahertz (THz) communications have emerged as a key technology for escalating data rates in future generation wireless networks.
However, severe propagation losses at THz frequencies pose significant challenges, which can be mitigated via ultra-massive multiple-input multiple-output (UM-MIMO) systems employing highly directional transmissions. 
To this end, codebook-based analog beamforming constitutes a realistic solution, eliminating the need for explicit channel estimation. However, in UM-MIMO systems, the use of extremely narrow beams makes beam training and alignment increasingly challenging, leading to a substantial increase in the number of codewords to be tested and, thus, to high computational complexity. In this paper, a novel artificial neural network architecture for low-complexity beam training in UM-MIMO THz systems is presented, which does not require a constant feedback link between transmitter and receiver to obtain the best beamformer and combiner pair. An inception and residual network, which is trained based on the received signal powers using the transmit and receive codewords generated from predefined hierarchical codebooks, is designed. Our numerical investigations demonstrate that the proposed machine learning approach significantly reduces the complexity of UM-MIMO transmit and receive beamforming design, as compared to the standard exhaustive and hierarchical beam searching methods.
\end{abstract}
\section{Introduction}
To meet the growing data requirements in future wireless networks, the terahertz (THz) frequency band emerges as a promising facilitator \cite{SarieddeenPoE21, PreethamIABFNWF}. This is due to the large available bandwidth at the THz frequency spectrum. Recent progress in semiconductor industries has facilitated the creation of more compact THz devices \cite{MatosICEAA24}, overcoming a traditional hurdle in deploying THz technology \cite{THzJSAC21}. 
Although relevant systems are still in an early stage of development, few standardization and regulatory processes have already begun \cite{ETSI}, \cite{THzUAVCommMag22}.

One of the critical challenges for THz wireless communications is the substantial propagation losses linked with operating at higher frequencies due to molecular absorption. To address this challenge, ultra-massive multiple-input-multiple-output (UM-MIMO) systems can be employed \cite{NingOJCS23, SheemarWCNC24}. For such systems, digital beamforming techniques are the most efficient in terms of performance and flexibility. However, the higher-cost architecture pertaining to costly radio-frequency (RF) chains and substantial energy requirements of mixed-signal components make it unaffordable for UM-MIMO THz systems. On the other hand, analog beamforming requires a single RF chain and is preferable for these systems from the perspective of cost and energy consumption. However, it offers less flexibility compared to digital beamforming. To strike a balance between cost and performance, hybrid beamforming, which is a combination of both analog and digital beamforming, is an emerging approach\cite{ElbirTVT23, HuangJSAC21, ShlezingerIWC21, GavriilidisICASSP24}.

However, it is noteworthy that digital and hybrid beamforming techniques require full channel state information (CSI) to enable the design of optimal transmit and receive beamformers and combiners, which becomes exceedingly complex and expensive for UM-MIMO. To this end, analog beamforming overcomes this problem and codebook-based beam training schemes can be leveraged \cite{XiaoTWC16, ZhangTC22, GeorgeSAM22, AlexandropoulosWCNC21} to realize analog beamforming without the need for CSI. Such approaches hold even greater promise for reducing complexity in UM-MIMO THz systems. Beam training identifies the most robust narrow beam pair from a predefined codebook consisting of various beamformers corresponding to different angles of arrival (AoA) and angles of departure (AoD), and subsequently utilizing this pair directly for data transmission. To obtain the optimal transmit and receive beamforming codewords, an exhaustive search can be employed, which sequentially tests all possible beam directions to find the best pair. Such a method is computationally expensive as well as time consuming, with complexity growing exponentially with the size of transmit/receive antennas. Hierarchical codebooks have been proposed to expedite the search process with significantly less complexity \cite{XiaoTWC16, GavriilidisTWC25} compared to the exhaustive search. However, this still remains a major challenge when the number of antenna elements is extremely large, which is the case of UM-MIMO THz systems. 
 Additionally, various learning-based techniques have been introduced for codebook learning \cite{zhangSPAWC20, ZhangTC22}, however, these methods are primarily driven by the high complexity associated with beam search.

Motivated by the above, in this paper, a novel neural network architecture for single-RF transmit and receive analog beamforming design in UM-MIMO THz systems is presented. The proposed architecture, an inception and residual network (Incept-ResNet), is trained offline using received power measurements derived from predefined hierarchical codebooks. Once trained for a specific deployment area, our learning framework can efficiently predict the optimal transmit beamformer, receive combiner, and the corresponding predicted received power. Notably, the proposed low-complexity beam training framework eliminates the need for continuous feedback between the transmitter and receiver, significantly reducing the beam alignment overhead. The presented results verify Incept-ResNet's reduced beamforming overhead as well as its superiority over the exhaustive and traditional hierarchical beam searching techniques. While prior learning-based works such as \cite{zhangSPAWC20, ZhangTC22} focus on the complex task of learning the entire beam codebook from scratch based on environmental data, our approach addresses a different but equally critical challenge. In particular, our approach reduces the beam search complexity within predefined hierarchical codebooks, eliminating feedback overhead and enabling low-complexity inference. The key innovation lies not in generating the codebook, but in creating an efficient inference model that predicts the optimal narrow beam pair by using only the power measurements from the initial, widest beams. This drastically reduces the search overhead from a multi-stage logarithmic search.
\section{System Model and Problem Formulation} \label{sec2}
We consider an outdoor point-to-point UM-MIMO THz communication system consisting of a transmitter (Tx) equipped with $N_{\rm T}$ antenna elements and a receiver having $N_{\rm R}$ antennas. Both Tx and Rx are capable of realizing highly directional analog beamforming with a single respective RF chain, ensuring low energy consumption and cost. Let $x$ denote the transmitted data symbol from Tx with unit power. The corresponding received signal at Rx can be expressed as
\begin{align}\label{y_rec}
y=\sqrt{P_t}{\bf{w}}_{\mathrm{R}}^{\mathrm{H}}{\bf{H}}{{\bf{w}}_{\mathrm{T}}} x+{\bf{w}}_{\mathrm{R}}^{\mathrm{H}}{\bf{n}}_r,
\end{align}
where ${\bf{w}}_{\mathrm{T}}\in \mathbb{C}^{N_{\rm T} \times 1}$ and ${\bf{w}}_{\mathrm{R}}\in \mathbb{C}^{N_{\rm R} \times 1}$ are the beamformer and combiner, respectively, with $P_t$ denoting the transmit power, and $(\cdot)^{\rm{H}}$ denotes the conjugate transpose operation. ${\bf{H}}\in \mathbb{C}^{N_{\rm R} \times N_{\rm T}}$ is the channel matrix and ${\bf{n}}_r \in \mathbb{C}^{N_{\rm R} \times 1}$ is the additive white Gaussian noise vector with zero mean and variance $N_0$.
\subsection{THz Channel Model}
Noting that THz channel responses are LoS-dominant in an outdoor scenario, we consider uniform linear arrays at Tx and Rx with the intra-element spacing of half the wavelength. Assuming communication in the far-field\footnote{Larger numbers of antennas can be placed in a small antenna aperture size, owing to the very small wavelengths at THz. The Rayleigh distance is defined as ${\rm{RD}}=(N_X-1)^2\lambda /2$, indicating that the boundary between the near- and far-fields regimes can be very small. It has been reported in \cite{NingOJCS23} that this distance for even 2500 antenna elements is only 1.2 meters.}, the UM-MIMO channel response is given as 
 ${\textbf{H}}=\sqrt{G N_{\rm R}N_{\rm T}}\psi{\textbf{a}}_{\rm R}(N_{\rm R},\theta_{\rm R}){\textbf{a}}_{\rm T}^H(N_{\rm T},\theta_{\rm T})\alpha(f_c,d_{t,r}),$
where $G=G_tG_r$ represents the Tx/Rx antenna gain, $\psi$ is a complex Gaussian distributed channel coefficient, and $\theta_{\rm T}$, $\theta_{\rm R}$ are the angle of departure (AoD) and angle of arrival (AoA), respectively. The vectors $\textbf{a}_{\rm R}(\cdot)$ and $\textbf{a}_{\rm T}(\cdot)$ denote receive and transmit array response given as 
 $\textbf{a}_X(N_X,\theta_X)=\frac{1}{\sqrt{N_X}}[1, e^{j\pi 1 \cos(\varphi)},\cdots,e^{j\pi (N-1) \cos(\varphi)}]^T,$
with $X\in \{\rm T, \rm R\}$. The scalar $\alpha(f_c,d_{t,r})$ denotes the path loss including both the free-space and molecular absorption losses. Specifically, the term $\alpha(f_c,d_{t,r})$ can be expressed as 
 $\alpha(f_c,d_{t,r})=\frac{c}{4\pi f_c d_{t,r}}e^{-\frac{1}{2}\kappa(f_c)d_{t,r}},$ 
where $c$, $f_c$, and $d_{t,r}$ represent the speed of light, the carrier frequency, and the distance between Tx and Rx, respectively \cite{ElbirTVT23}. The term $\kappa(f_c)$ denotes the frequency-dependent medium absorption coefficient.
\subsection{Beam Training Problem}
\begin{figure}
    \centering    \includegraphics[width=7 cm,height=3.2cm]{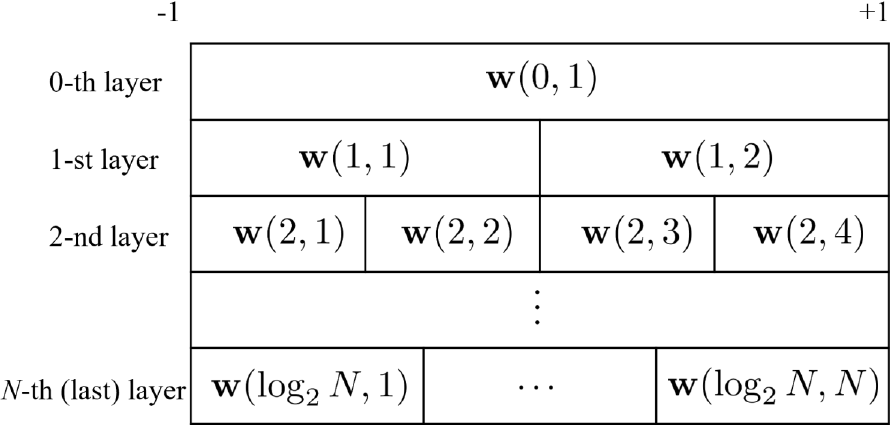}
    \caption{Typical structure of a 2-tree (binary-tree) codebook with $N_{\rm T}=N_{\rm R}=N$.}
    \label{fig:codebook_strct}
\end{figure}

For the considered UM-MIMO THz system, the beamforming problem can be posed as 
\begin{align}
(\mathbfcal{P}):\,&\max_{{\bf{w}}_{\mathrm{T}},{\bf{w}}_{\mathrm{R}}} \, \gamma|{\bf{w}}_{\mathrm{R}}^{\mathrm{H}}\bf{H}{\bf{w}}_{\mathrm{T}}|^2 \\
\mathrm{subject\, to :}\,& ||{\bf{w}}_{\mathrm{T}}||^2 = ||{\bf{w}}_{\mathrm{R}}||^2 =1,
\end{align}
where $\gamma=P_t/N_0$ is the transmit signal-to-noise ratio (SNR). 
Evidently, if the CSI is known, the optimal beamformer (${\bf{w}}_{\mathrm{T}}^*$) and combiner  $({\bf{w}}_{\mathrm{R}}^*)$ can be obtained from $\mathbf{H}$'s singular value decomposition \cite{NingOJCS23}. However, CSI acquisition in UM-MIMO THz systems is highly complex. To this end, we employ codebook-based beamforming, which can overcome this bottleneck \cite{XiaoTWC16}. As the resulting beamforming design is independent of SNR, $(\mathbfcal{P})$ can be recast as
\begin{align}
(\mathbfcal{P}_1):\,&\max_{{\bf{w}}_{\mathrm{T}},{\bf{w}}_{\mathrm{R}}} \, |{\bf{w}}_{\mathrm{R}}^{\mathrm{H}}\bf{H}{\bf{w}}_{\mathrm{T}}|^2 \\
\mathrm{subject\, to :}\,& {\bf{w}}_{\mathrm{T}} \in \mathcal{W}_{\mathrm{T}}, {\bf{w}}_{\mathrm{R}} \in \mathcal{W}_{\mathrm{R}},
\end{align}
where $\mathcal{W}_{\mathrm{T}}$ ($\mathcal{W}_{\mathrm{R}})$ are the predefined Tx (Rx) codebooks comprising several antenna response vectors forming beams in different directions based on the AoAs/AoDs. For instance, the codebook can be a classical discrete Fourier Transform (DFT) codebook wherein the codewords form single lobe beams in the whole angular space \cite{ZhilinTVT22}. However, for such codebooks, the exhaustive search is employed to find the optimal transmit/receive codewords, which incurs significant complexity in UM-MIMO. Hierarchical codebooks offer a more practical approach that demands less computational complexity, utilizing a layered search that begins with wider beams and progressively refines the search to identify optimal narrow beams (e.g., \cite{GeorgeSAM22}). 
In our work, we follow the antenna element deactivation codebook design approach in \cite{XiaoTWC16}, according to which the codewords can be given by
\begin{align}
    \mathbf{w}(k, n)\!=\!\left[\mathbf{a}_X\!\left(2^k,-1\!+\!\frac{2 n-1}{2^k}\right)^{\mathrm{T}}, \mathbf{0}_{\left(N_X-2^k\right) \times 1}^{\mathrm{T}}\right]^{\mathrm{T}},
\end{align}
where $n$ denotes the $n$-th codeword in $k$-th layer and $(\cdot)^{\rm{T}}$ is the transpose operation.
A typical structure of a binary-tree hierarchical codebook is shown in Fig.~\ref{fig:codebook_strct}. 
Note that the computational complexity of such hierarchical
codebooks increases with the dimension of the system, thus, it can be very challenging at THz with UM-MIMO arrays.
    
\begin{figure*}[!ht]
    \centering
    \centerline{\includegraphics[width=5.5 in, height=2.0in]{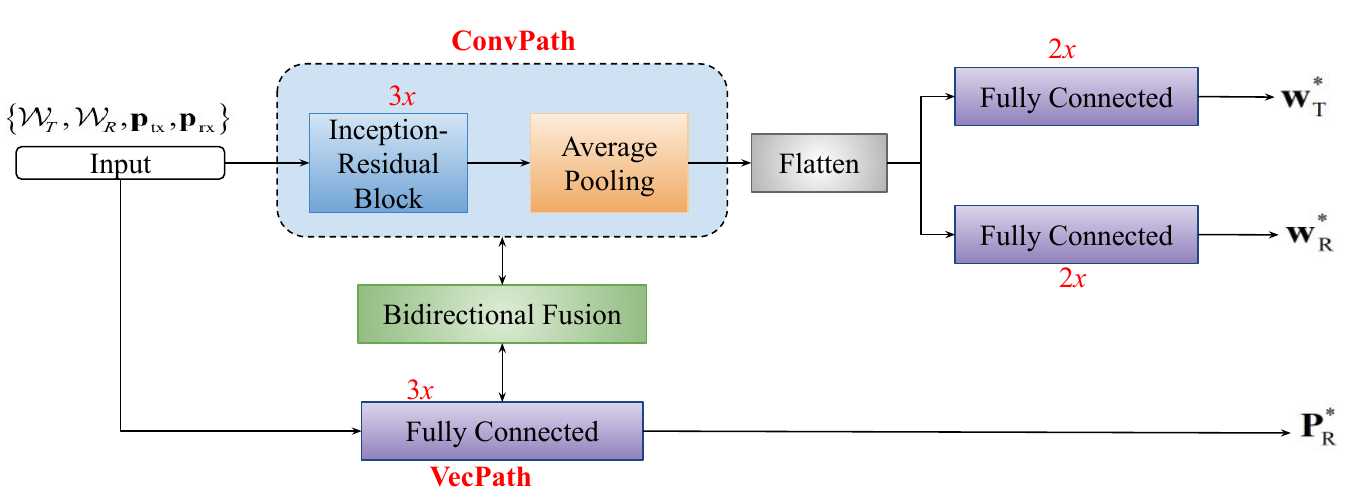}}
    \caption{Architecture of the proposed Incept-ResNet for beam training.}
    \label{fig:model}
\end{figure*}

\section{Proposed Beam Training Approach} \label{sec3}
In this section, we present a novel learning-based approach for solving problem ($\mathbfcal{P}_1$) with low complexity. We commence by training a neural network utilizing the aforedescribed hierarchical codebook. After training, the proposed neural network can discern the optimal beamformer and combiner with reduced computational requirements in contrast to conventional hierarchical search methods. It is worth noting that, while several learning-based approaches have been proposed for learning codebooks \cite{zhangSPAWC20,ZhangTC22}, they primarily address the high complexity associated with beam searching. In contrast, the objective of this work is to reduce the search complexity involved in identifying the best narrow beam pair using a hierarchical codebook. 

\subsection{Data Generation and Training}
To generate the necessary training dataset for the proposed neural network, we consider a coverage region where the Tx is located at the center of a circle with a radius $R$. For each random location of the Rx, the received powers, i.e., $\gamma{|{\bf{w}}_{\mathrm{R}}^{\mathrm{H}}\bf{H}{\bf{w}}_{\mathrm{T}}|}^2$, at the Tx and the Rx are calculated using the predefined hierarchical codebooks $\mathcal{W}_{\mathrm{T}}$ and $\mathcal{W}_{\mathrm{R}}$.
To obtain the received powers, first, the Tx is fixed in an omnidirectional mode with the codeword $\textbf{w}_{\rm T}=\textbf{w}(0,1)$ from the $0$-th layer of the codebook. Next, an $M$-way tree search is performed at the Rx for $\log_M(N_{\rm{R}})$ stages to obtain the best Rx's codeword $\textbf{w}^*_{\rm{R}}$, and then the power measurements are stored in a vector $\textbf{p}_{\rm rx}$ of dimension $M\log_M(N_{\rm{R}})\times 1$. The Rx's codeword $\textbf{w}_{\rm{R}}$ is now set in a directional mode 
with the codeword $\textbf{w}^*_{\rm{R}}$ obtained in the preceding step. Subsequently, an $M$-tree search is performed at the Tx's codebook $\mathcal{W}_{\mathrm{T}}$ for $\log_M(N_{\rm{T}})$ stages to obtain the best Tx's codeword $\textbf{w}^*_{\rm T}$, and the power measurements are stored in a vector $\textbf{p}_{\rm tx}$ of dimension $M\log_M(N_{\rm{T}})\times 1$.

The role of the proposed neural network is to take the Rx power vectors along with the codebooks $\mathcal{W}_{\mathrm{T}}$, $\mathcal{W}_{\mathrm{R}}$, and return the optimal codewords $\textbf{w}^*_{\rm T}, \textbf{w}^*_{\rm R}$, as shown in Fig.~\ref{fig:model}. During the training, the mean square error (MSE) is used as the loss function.
Once the network is trained within the area of interest, it can efficiently obtain during the inference/prediction
stage the best Tx beamformer,
Rx combiner, and corresponding optimal received power with low complexity.
Although hierarchical search is used to generate training labels, this is performed offline solely for dataset construction. During inference, the proposed model does not rely on hierarchical search and directly predicts the beam pair from first-layer measurements.
\subsection{Proposed Neural Network Design}
In this section, we present the proposed Incept-ResNet architecture, which is designed to enable efficient beam training. The architecture comprises two parallel processing pipelines: \textit{(i)} ConvPath, which integrates the principles of inception and residual networks to extract spatial features; and \textit{(ii)} VecPath, which consists of fully-connected (FC) layers with batch normalization (BN) to process vectorized inputs. These parts interact through a feature fusion mechanism to enhance learning. Incept-ResNet design and functional components are summarized as follows.
\begin{itemize}
    \item Inception Module: 
    This module is a block that applies several convolutions of different sizes to the input and then concatenates the results. This allows the network to capture information at various scales. Let $X$ be the input to Incept-ResNet including Tx and Rx
    codebooks $\mathcal{W}_{\mathrm{T}}$ and $\mathcal{W}_{\mathrm{R}}$  and their corresponding powers $(p_1^{\rm{T}},p_2^{\rm{T}})$ and $(p_1^{\mathop{\rm R}\nolimits}, p_2^{\rm{R}}$); $(p_1^{\rm{T}}, p_2^{\rm{T}})$ and $(p_1^{\mathop{\rm R}\nolimits}, p_2^{\rm{R}})$ are the powers corresponding to the first layer of the codebook at the Tx and the Rx, respectively. The output of the Inception module can be described as ${Y^{{\text{Incept}}}} = \left[ {{f_1}\left( X \right);{f_2}\left( X \right);{f_3}\left( X \right);{f_4}\left( X \right)} \right],$
      where ${f_1}\left( X \right),{f_2}\left( X \right),{f_3}\left( X \right)$, and ${f_4}\left( X \right)$ are convolution operations with different filter sizes.
      \item Residual Connection: Let  $F\left( {{Y^{{\text{Incept}}}}} \right)$ represent the output from a stack of layers applied to ${Y^{{\text{Incept}}}}$. Then the output of a residual block is given by ${Y^{{\rm{Res}}}} = F\left( {{Y^{{\rm{Incept}}}}} \right) + {Y^{{\rm{Incept}}}}.$
      Such residual connections are particularly effective in addressing the vanishing gradient problem, which may arise when very small or near-zero gradients diminish progressively across layers in deep neural networks. By enabling direct gradient flow, the proposed architecture preserves learning efficiency in deeper networks.
\end{itemize}
After each convolution in the Inception module and after the residual summation, BN and rectified linear unit (ReLU) activation function are applied. 
To form the proposed Incept-ResNet, multiple inception and residual blocks are stacked together. Moreover, max pooling is applied within the Inception modules, while average pooling (AvP) is applied after the stacked inception and residual blocks to reduce the spatial dimension of the feature maps. Finally, flattening and FC layers are stacked to convey the output codebooks. 

Fusing multilevel features in deep learning is a powerful technique that significantly enhances the performance of predictive models. This approach leverages the hierarchical nature of deep learning architectures, where different layers capture information at various levels of abstraction. Motivated by these reasons, we design a bidirectional fusion block with convolutional layers, and linear layers to combine features from different levels. This block helps in fusing and transferring features between ConvPath and VecPath. This can significantly improve the model's ability to understand the context for prediction.
The outputs $Y$ are obtained at the end of the two pipelines and include the beamformer ${\bf{w}}_{\rm{T}}^*$, combiner ${\bf{w}}_{\rm{R}}^*$, and the corresponding average received power $\textbf{P}_{\rm{R}}^*$. 
\subsection{Inference}
In the inference/prediction stage, the neural network requires the received power inputs corresponding to the first layer of the hierarchical codebook (wider beams) at both the Tx and the Rx ends. Specifically, for the inference, the neural network just needs the power measurements corresponding to the first layers of codebooks $\mathcal{W}_{\mathrm{T}}$, i.e., $p_1^{\rm{T}},p_2^{\rm{T}}$, and $\mathcal{W}_{\mathrm{R}}$, i.e., $p_1^{\mathop{\rm R}\nolimits}, p_2^{\rm{R}}$, to obtain the optimal $\textbf{w}^*_{\rm T}$ and $\textbf{w}^*_{\rm R}$. Based on this, the network also predicts the average received power $({|{\bf{w}}_{\mathrm{R}}^{*\mathrm{H}}\bf{H}{\bf{w}}^*_{\mathrm{T}}|}^2)$ at the Rx at any given location.
This is in contrast to the conventional hierarchical search methods, where the search is performed until the last layer of the codebook corresponding to narrow beams to obtain the optimal codewords. Also, our scheme does not require iterative multi-stage feedback or continuous index exchange during beam training. 
In practice, the proposed framework assumes that the receiver can convey a compact vector of first-layer power measurements (or a compressed representation thereof) to the transmitter once per coherence interval. Unlike conventional hierarchical search, the proposed method does not require iterative, stage-wise beam index exchange during the beam training process.
Consequently, the proposed learning-based Tx/Rx beamforming can reduce the total computations from  $2M\log_M N$ in the case of conventional hierarchical search to $2M$, thereby significantly reducing the complexity. \\
\vspace{-0.1in}
\subsection{Complexity and Convergence }
The computational complexity of the proposed learning-based approach can be divided into training complexity and inference complexity. To calculate the training complexity, we consider the number of floating-point operations (FLOPS) in each block. The dimension of the input codebook is ${N_{\rm T}} \times {N_{\rm T}} \times {S_{\rm T}}$ and ${N_{\rm R}} \times {N_{\rm R}} \times {S_{\rm R}}$ with ${S_{\rm T}}$ and ${S_{\rm R}}$ respectively represent the depth of the input codebooks in the Tx and the Rx. The dimensions of the output codebooks are given by ${N_{\rm T}} \times 1$ and ${N_{\rm R}} \times 1$. Since other inputs are either short vectors or singular values, their impact is negligible in the overall scale of the problem. The asymptotic complexity of the ConvPath is ${{\cal O}^{{\rm{incep}}}} = {\cal O}\left( {3{K^2}{N^2}S} \right)$, where $K$ represent the number of kernels, $N = {N_t} = {N_r}$, and $S = {S_t} = {S_r}$. On the other hand, we have parallel VecPath asymptotic complexity ${{\cal O}^{{\rm{ff}}}} = 2 \times \sum\limits_{l = 1}^L {{n_{l - 1}} \times {n_l}} $, where $n$ and $L$ represent the number of neurons and layers, respectively. The bidirectional fusion has complexity  ${O^{{\rm{bi}}}} = O\left( {{H_l}{W_l}{K_l}{C_l} + 2n + 1} \right)$, where ${H_l},{W_l},$ and ${C_l}$ represent the height, width, and depth of the feature representation of layer $l$. Therefore, the overall training complexity in the worst case can be given by ${\cal O} = {{\cal O}^{{\rm{incep}}}} + {{\cal O}^{{\rm{ff}}}} + {O^{{\rm{bi}}}}$. During the inference, the training complexity has no impact on the inference speed of the model. Hence, the complexity scales quadratically with the input dimension and is influenced by the design parameters of the inception blocks as well as the dimensions in the VecPath. 
\begin{figure}
    \centering
    \includegraphics[width=8.4cm,height=5.4cm]{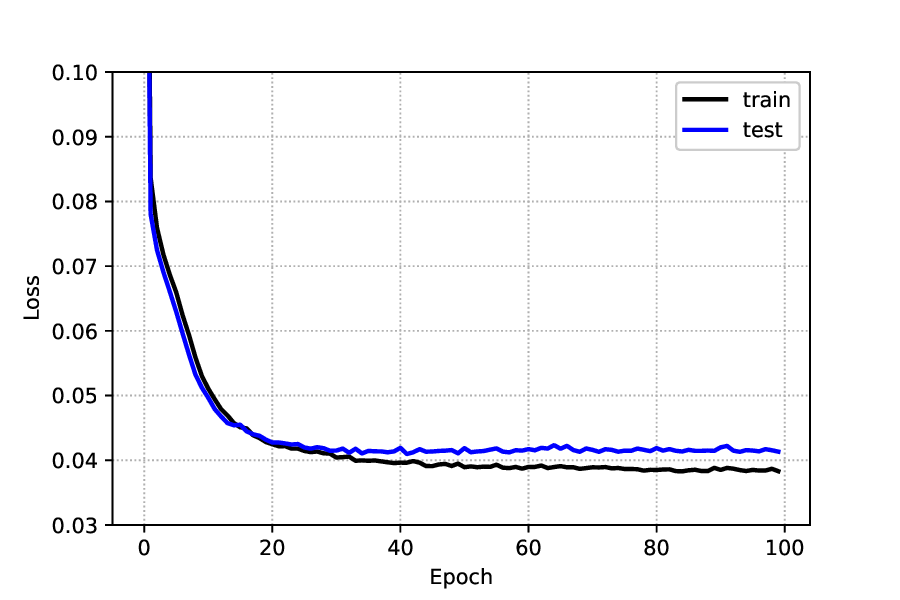}
    \caption{Training and testing loss of the proposed Incept-ResNet model.}
    \label{fig:loss}
\end{figure}

Fig. \ref{fig:loss} illustrates the convergence behavior of the model for different learning rates. It can be seen that, When the learning rate is set to $\delta = 0.001$, the model converges more rapidly, achieving stable performance within approximately 40 epochs. In contrast, a higher learning rate of $\delta = 0.003$ results in slower convergence, requiring around 70 epochs to stabilize. Therefore, $\delta = 0.001$ is the default for the result in Fig. \ref{fig:loss} as well as the other experimental results. The relatively small gap between training and testing loss indicates good generalization capability. Notably, in the early stages (as shown by the black line), the training performance initially lags slightly behind the test performance, which is common in the early phase of training. As training progresses, the model achieves stable and strong generalization performance.

\section{Numerical Results} \label{sec4}
In this section, we present performance evaluation results of our proposed method against the baseline approaches to assess its effectiveness. The operating frequency was set to $f_c=0.14$ THz and the absorption coefficient as $\kappa(f_c)=6\times10^{-5}/m$. Antenna gains were set as $G_t=G_r=4+10\log_{10}(\sqrt{N_X})$ \cite{NingTVT21}. We have considered $N_{\rm T}=N_{\rm R}=N$, $M=2$, and $\gamma =0 {\, \rm dB}$. The parameters for designing the neural network are given in Table \ref{table1}. Our neural network framework was implemented in PyTorch 2.0 and Python 3.8 platforms. 
The dataset consisted of 10000 training samples and 2000 test samples. To avoid getting smaller gradient descent during training, the bounded logarithmic normalization method was used to normalize the power \cite{Adam98575}.
\begin{table}[]
\centering
\caption{Incept-ResNet parameters.}
\label{table1}
\begin{tabular}{|l|l|}
\hline
\textbf{Parameter} & \textbf{Value} \\ \hline
\multicolumn{2}{|c|}{\textbf{ConvPath}} \\ \hline
Number of inception blocks & 3 \\ \hline
Kernel sizes in each block & 1x1, 1x3, 3x1, 3x3 \\ \hline
Stride & 2 \\ \hline
Pooling in each block & 2x2 max pooling \\ \hline
Pooling & Average pooling \\ \hline
Activation function & ReLU \\ \hline
\multicolumn{2}{|c|}{\textbf{VecPath}} \\ \hline
Number of linear layers & 4 \\ \hline
Activation function & ReLU \\ \hline
\multicolumn{2}{|c|}{\textbf{Bidirectional Fusion}} \\ \hline
Kernel sizes & 1x1, 3x3, 3x3 depthwise \\ \hline
stride & 1, 2 \\ \hline
Linear layers & 2 \\ \hline
Flattening    & 1 \\ \hline
\multicolumn{2}{|c|}{\textbf{Training Parameters}} \\ \hline
Learning rate & 0.001 \\ \hline
Batch size	&32\\ \hline
Dropout &	0.5\\ \hline
Optimizer 	&Adam\\ \hline
Loss function&	Customized MSE\\ \hline
\end{tabular}
\end{table}

Table \ref{table2} compares the computational complexity of the proposed method with that of baseline beam training protocols \cite{NingOJCS23}. It can be observed that our learning-based approach exhibits significantly lower inference complexity compared to conventional methods. 
Note that the training complexity was excluded from this comparison, since training is conducted offline prior to deployment in a given environment. Once trained, the model enables efficient selection of the best beamformer and combiner with minimal computational overhead.
\begin{table}[t]
\centering
\caption{Complexity comparison of different beamforming schemes.}
\label{table2}
\begin{tabular}{|l|l|}
\hline
\textbf{Training Protocol}                      & {\color[HTML]{333333} \textbf{Complexity}} \\ \hline
Exhaustive Search                               & $N^2$                                      \\ \hline
One-side Search                                 & $2N$                                         \\ \hline
Adaptive Search                                 & $4\log_2N$                                 \\ \hline
Parallel Search                                 & $N^2/N_{RF}$                               \\ \hline
One-side M-Tree Search                          & $2M\log_M N$                               \\ \hline
Both-side M-Tree Search & $M^2\log_M N$                              \\ \hline
Proposed NN-based Method (ex. Training)         & $2M$                                       \\ \hline
\end{tabular}
\end{table}

\begin{figure}
    \centering
    \includegraphics[width=8cm,height=6cm]{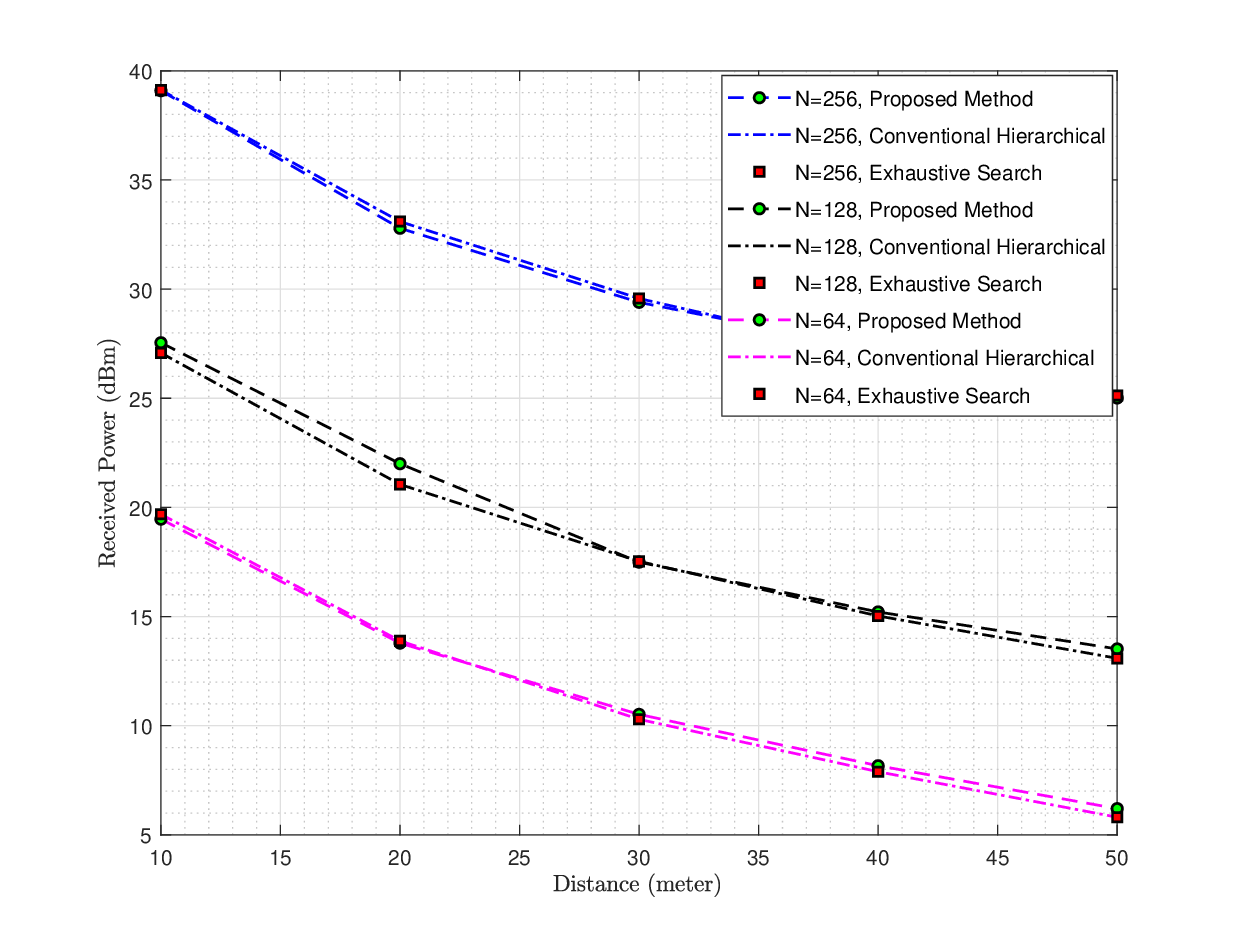}
    \caption{Average power versus distance.}
    \label{fig:avp}
\end{figure}

Fig.~\ref{fig:avp} compares the performance of the proposed approach with exhaustive search (complexity $\mathcal{O}(N^2)$) and conventional hierarchical search (complexity $\mathcal{O}(2M \log_M N)$) by plotting the average received power as a function of the distance between the Tx and Rx. Experiments are conducted for different antenna array sizes at both Tx and Rx, specifically $N = 256$, $128$, and $64$. As observed, the proposed method achieves performance comparable to the baseline schemes, while incurring significantly lower complexity of only $\mathcal{O}(2M)$. This validates the effectiveness of the proposed strategy. Additionally, as expected, the average received power decreases with increasing distance. Furthermore, although the system is evaluated with up to 256 antennas, the proposed scheme is inherently scalable and well-suited for deployment in systems with larger antenna arrays.

\begin{figure}
    \centering
    \includegraphics[width=8cm,height=6cm]{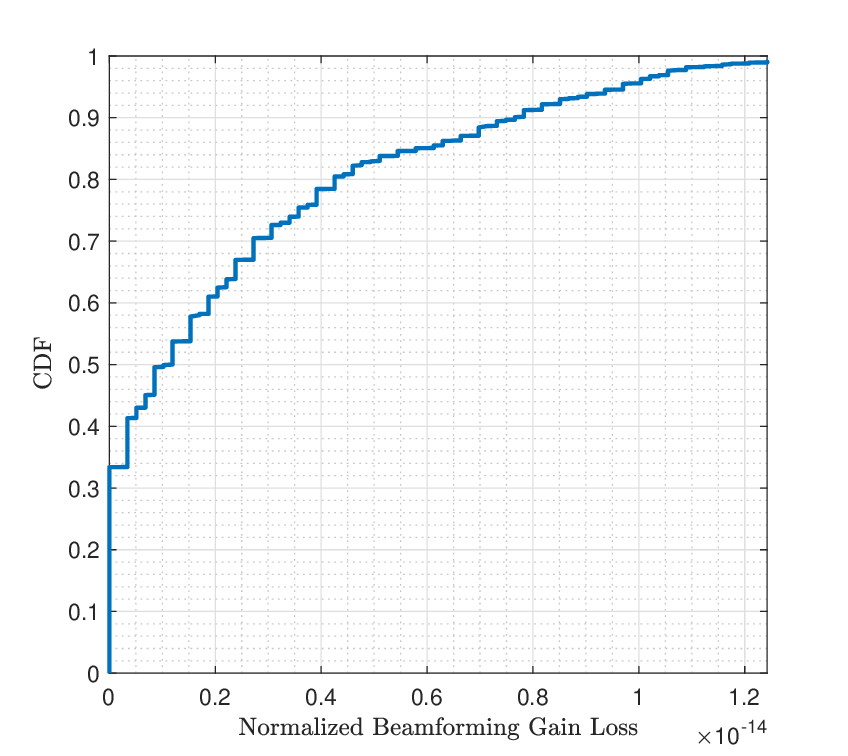}
    \caption{CDF of beamforming gain loss.}
    \label{fig:cdf}
\end{figure}
To assess the robustness of the proposed beam training framework, we examine the cumulative distribution function (CDF) of the beamforming gain loss relative to exhaustive search. Let $P_{\mathrm{exh}} = \bigl| \mathbf{w}_R^{\mathrm{exh}\,H} \mathbf{H} \mathbf{w}_T^{\mathrm{exh}} \bigr|^2$
and $P_{\mathrm{prop}} = \bigl| \mathbf{w}_R^{\mathrm{prop}\,H} \mathbf{H} \mathbf{w}_T^{\mathrm{prop}} \bigr|^2$
denote the received beamforming gains obtained via exhaustive search and the proposed method, respectively. The normalized beamforming gain loss is defined as $\Delta_{\mathrm{norm}} \triangleq 
\frac{P_{\mathrm{exh}} - P_{\mathrm{prop}}}{P_{\mathrm{exh}}}.$
Using multiple independent channel realizations, the empirical CDF of the gain loss is given by $F_{\Delta}(x) = \Pr\!\left( \Delta_{\mathrm{norm}} \le x \right).$
Fig.~\ref{fig:cdf} illustrates the CDF of the normalized beamforming gain loss relative to exhaustive search. It is observed that the proposed method incurs negligible loss which confirms the near-optimal performance of the proposed framework
\section{Conclusion} \label{sec5}
This paper proposed a novel neural network architecture for low-complexity beamforming in UM-MIMO THz communication systems. The developed Incept-ResNet was trained using predefined hierarchical codebooks and relied solely on power measurements. This feature eliminates the need for CSI acquisition, which is especially advantageous for THz communication
deployment. The conducted performance evaluation demonstrated that Incept-ResNet significantly reduces the complexity of obtaining the
best beamformer and combiner pair, as compared to exhaustive and classical hierarchical search methods. 

\section*{Acknowledgement}
 This work was partially supported by the Telecom Technology Development Fund (TTDF) under the project ID TTDF$/\text{6G}/$461, approved as
part of the TTDF scheme of the Digital Bharat Nidhi (DBN), Department of Telecommunications (DoT). The work is also supported in part by the SNS JU project TERRAMETA under the European Union's Horizon Europe research and innovation programme under Grant Agreement No 101097101, including top-up funding by UKRI under the UK government’s Horizon Europe funding guarantee.
\bibliographystyle{IEEEtran}
 \bibliography{bibl}
 \balance
\end{document}